\documentclass[reprint,
 superscriptaddress,
 amsmath,
 amssymb,
 aps,
 pra,
]{revtex4-1}
\usepackage[utf8]{inputenc}
\usepackage{graphicx}
\usepackage{dcolumn}
\usepackage{bm}
\usepackage{natbib}
\setcitestyle{super}
\usepackage{dirtytalk}
\usepackage{csquotes}
\usepackage{subcaption}
\usepackage{multirow}
\usepackage{textgreek} 

\begin{document}

\title{Jamming-controlled stochasticity in metal-insulator switching}
\author{Nicolò D'Anna}
\email{ndanna@ucsd.edu}
\affiliation{University of California San Diego, La Jolla, CA 92093, USA}
\author{Nareg Ghazikhanian}
\affiliation{University of California San Diego, La Jolla, CA 92093, USA}
\author{Katherine Matthews}
\affiliation{University of California San Diego, La Jolla, CA 92093, USA}
\author{Daseul Ham}
\affiliation{Pohang Light Source-II (PLS-II) Beamline Department, Pohang Accelerator Laboratory, POSTECH, Pohang 37673, Republic of Korea}
\author{Su Yong Lee}
\affiliation{Pohang Light Source-II (PLS-II) Beamline Department, Pohang Accelerator Laboratory, POSTECH, Pohang 37673, Republic of Korea}
\author{Alex Frano}
\affiliation{University of California San Diego, La Jolla, CA 92093, USA}
\author{Ivan K. Schuller}
\affiliation{University of California San Diego, La Jolla, CA 92093, USA}
\author{Oleg Shpyrko}
\affiliation{University of California San Diego, La Jolla, CA 92093, USA}

\begin{abstract}
Understanding and controlling phase transitions is a fundamental part of physics and has been central to many technological revolutions, from steam engines to field-effect transistors.
At present, there is strong interest in materials with strongly coupled structural and electronic phase transitions, which hold promise for energy-efficient technologies.
Utilizing a structural phase transition and controlling its plasticity naturally leads to built-in memory, a key feature for emulating neurons and synapses in neuromorphic technologies.
Here, \textit{operando} Bragg X-ray photon correlation spectroscopy is used to study the evolution of the nano-domain distribution at the micron-scale in neuromorphic devices made from the archetypal Mott insulator vanadium dioxide.
It is found that after electrical switching, slow nano-domain reconfiguration occurs on timescales of thousands of seconds and that the domains undergo a jamming transition, offering control over switching stochasticity at the micron scale. 
More precisely, repetitive above-threshold currents plastically drive the system into a jammed/glassy state where switching becomes deterministic, while sub-threshold currents erase the short-term memory contained in the nano-domain distribution, recovering stochastic switching, thus offering a path for in-device learning.
The results illustrate the importance of studying the nanoscale physics associated with phase transitions in strongly correlated materials, even for macroscopic devices, and offer guidance for future device operation schemes.
\end{abstract}
\maketitle

Strongly correlated "quantum" materials in which a structural and concomitant electronic phase transition can be controlled via electrical gates are promising candidates for future technologies\cite{hoffmann2022quantum,markovic2020physics}.
Of particular interest for energy-efficient neuromorphic technologies are materials where a percolating conductive filament embedded in an insulating matrix forms upon electrical switching\cite{hoffmann2022quantum,zhang2020understanding,li2018review,del2018challenges,markovic2020physics,sawa2008resistive}.
While electrical measurements are mostly sensitive to ensemble averages, over the typical mesoscopic scale of devices, the exact nanoscale domain distribution in and around the filament may determine electrical switching stochasticity and reproducibility\cite{del2018challenges,ghazikhanian2025,del2017electrically}. 
Thus, it is necessary to study the evolution of the microscopic phase distribution at the nanoscale across entire devices, as a function of time and switching state. Of particular interest is the extent the system retains memory of past states and whether these states can be erased or plastically modified\cite{fried2024film,ramirez2009first,vardi2017ramp}.

\begin{figure}[t] 
\centering   
    \includegraphics[width=\linewidth]{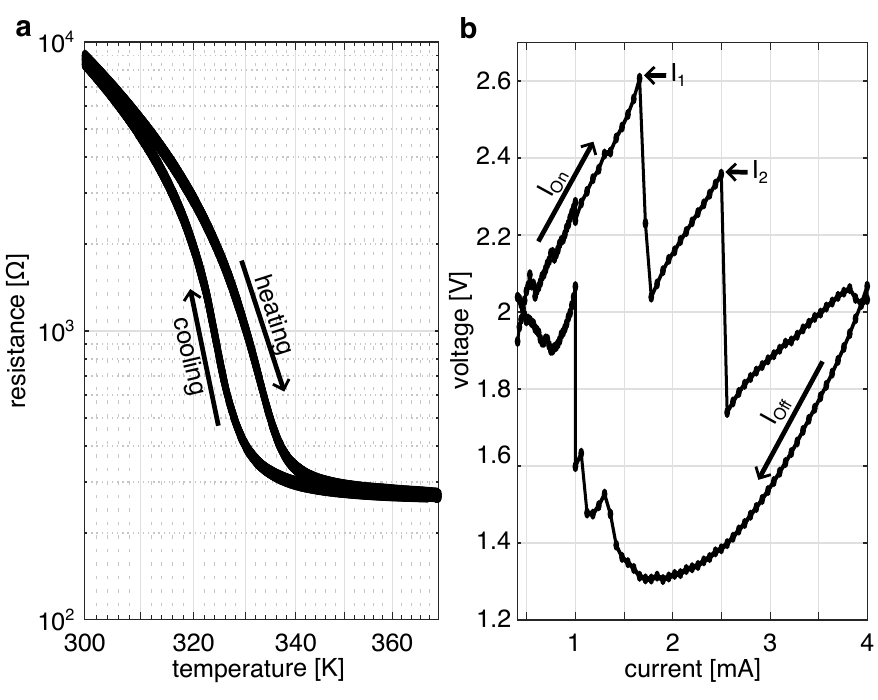} 
\caption{\textbf{Resistance vs temperature and current.} 
\textbf{a}, thermal switching measured prior to the XPCS experiment for the device exposed to a 0.1~pC/$\mu$m$^{2}$ Ga$^+$ dose.
\textbf{b}, electrical switching measured during the XPCS beamtime. $I_1$ and $I_2$ indicate the first and second switching currents.
}
\label{fig_RT}
\end{figure}

Here, \textit{operando} coherent Bragg X-ray diffraction\cite{shpyrko2014x} is utilized to track changes in the nanoscale distribution of the insulating 200$_{M1}$ and metallic 011$_R$ domains throughout micron-sized vanadium dioxide (VO$_2$) neuromorphic devices over periods of hours during electrical switching. By synchronously switching the devices and recording the diffraction, microscopic phase-distribution reproducibility is studied as a function of switching cycle.
Results show that prior to electrical switching, there is a stable domain distribution, which can be erased with sub-threshold currents (no percolating conductive filament, see Methods section and Fig.~\ref{fig_RT}), whereupon correlation with the initial distribution is lost after thousands of seconds; \textit{i.e.}, there is erasable short-term memory.
Additionally, slow domain reconfiguration is observed around the conductive filament while keeping the devices switched on with a constant current. 
Similarities to glass and jamming transitions, typical of soft matter and granular systems\cite{liu2010jamming,berthier2011theoretical,sillescu1999heterogeneity,mari2009jamming,liu2001jamming}, are found in the domain dynamics.
The associated dynamics undergo a jamming transition from liquid-like to glassy on timescales of thousands of seconds after switching.
Finally, it is found that driving the system into the jammed state by repetitively switching devices for hundreds of cycles, leads to a gradual transition from stochastic to deterministic filament formation. This demonstrates that the material learns and retains memory of the previously switched state.

\begin{figure*} 
\centering   
    \includegraphics[width=\linewidth]{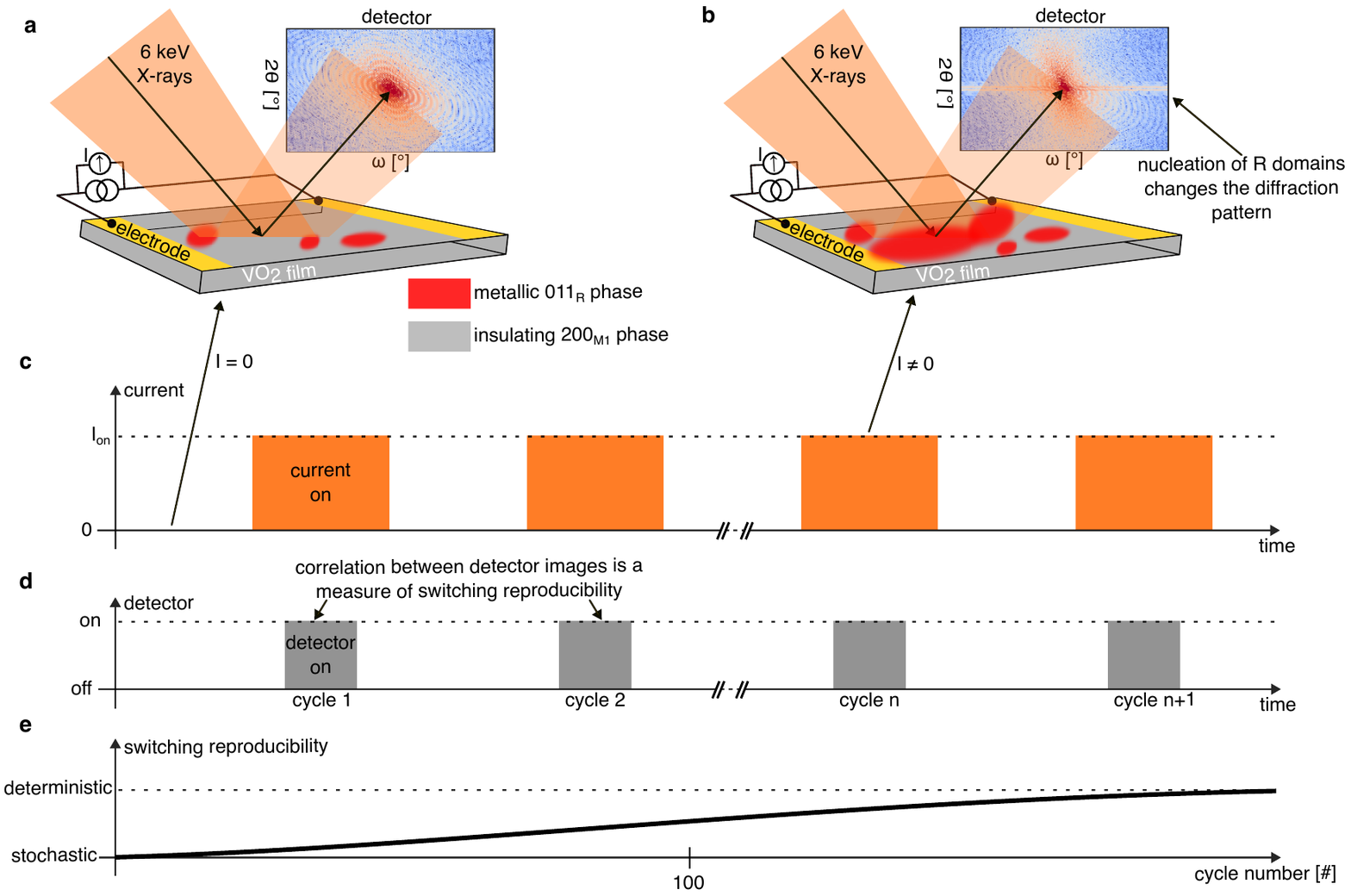} 
\caption{\textbf{\mbox{X-ray} diffraction and electrical setup.} 
    \textbf{a} and \textbf{b}, schematic representation of the X-ray photon correlation spectroscopy setup for a device in the Off and On states, respectively. The red area represents metallic domains within the insulating film (gray). Diffraction patterns are simulated based on the illustrated domain configurations. A current source connected to Ti/Au electrodes is used for electrical switching.
    \textbf{c}, experimental procedure for measuring switching reproducibility. The pulse sequences show the timing of electrical switching (top) and synchronized diffraction measurement (center). The bottom panel illustrates the transition from stochastic to deterministic switching observed in Fig.~\ref{fig_switch}.
    }
\label{fig_setup}
\end{figure*} 

\noindent\textbf{Vanadium dioxide neuromorphic devices}\\
VO$_2$ is an archetypal strongly correlated material exhibiting a metal-to-insulator transition (MIT) strongly coupled to a structural transition\cite{zylbersztejn1975metal,hu2023vanadium}.
The coupled electronic and structural phases have enabled nanoscale X-ray imaging of thermally\cite{d2025self,shao2025x} and electrically\cite{shabalin2020nanoscale} driven transitions, revealing the coexistence of conductive and insulating nanometer-scale domains.
Electrical switching of VO$_2$ devices results in the formation of a percolating metallic filament, a feature making VO$_2$ a strong candidate for neuromorphic applications\cite{mcwhan1970metal,duchene1971filamentary,zylbersztejn1975metal,mcwhan1973metal,adda2022direct,luibrand2023characteristic,rischau2024resistive}.
Full filament formation occurs on timescales of up to microseconds\cite{del2021spatiotemporal} and can be as fast as tens of picoseconds\cite{xu2023transient}. However, relaxation to the insulating state may take up to hours for large switching currents\cite{shabalin2020nanoscale}.
A method to control filament location and increase cycling stability is to locally induce defects through Ga or He focused ion beam irradiation of the VO$_2$ thin film\cite{xiang2021applications,ghazikhanian2023resistive,hofsass2011tuning,ghazikhanian2025electrically}. The defects reduce the insulating state resistivity, thus acting as a guide for filament formation\cite{ghazikhanian2023resistive}.
Two 5$\times$5~$\mu$m$^2$ VO$_2$ devices are studied in this work. They are defined by gold contacts (see Methods for detailed sample preparation). One device has a central 5$\times$5~$\mu$m region irradiated with a 0.1~pC/$\mu$m$^2$ Ga dose ($6.24 \times 10^{13}$~Ga ions/cm$^2$), the other is pristine. All electrical switching data shown in this article are from the irradiated device. Equivalent results were obtained from the pristine device (Supplemental Information), including the transition from stochastic to deterministic filament formation.

\noindent\textbf{Bragg X-ray photon correlation spectroscopy as a probe of microscale device stability}\\
X-ray photon correlation spectroscopy (XPCS) is a technique sensitive to all nanoscale domains within a micron-scale beam area\cite{shpyrko2014x}. It is uniquely adapted to studying the microscopic stability of electrically switched phase-change oxides, such as VO$_2$, over long periods of time (hours).
XPCS consists of recording, on a 2D detector, the diffraction from a sample as a function of time, illustrated in Fig.~\ref{fig_setup}a and~\ref{fig_setup}b. Because the X-rays are coherent, the diffraction images encode the full real-space distribution information in the interference speckle pattern. They represent the absolute value of the Fourier transform of the material's scattering length density profile.
On a Bragg peak, the diffraction stems from specific phases, in our case the insulating 200$_{M1}$ and metallic 011$_R$ phases of VO$_2$, Bragg angle $\sim$50$^\circ$ at 6~keV. Therefore, in this work the diffraction pattern characterizes the microscopic distribution of the metallic and insulating domains within the X-ray beam.

While real-space information is not obtained with XPCS, correlation between recorded diffraction images quantifies the amount of change as a function of time. If no change occurs, each subsequent diffraction image is perfectly correlated. 
Timescales and types of dynamics as well as domain speeds are obtained from the correlation decay, as described in the Methods section.
Note that the maximum value of the calculated correlation depends on the experimental setup geometry, stability, and the X-ray coherence length\cite{falus2006optimizing}. Thus, timescales associated with correlation decay are meaningful rather than absolute correlation values.
In this experiment, the X-ray beam is kept as large as the measured devices to capture the global phase distribution.
XPCS was measured while applying a constant current to characterize slow dynamics of the nano-domain distribution in and around the conductive filament under constant conditions.
Additionally, a novel approach was developed to study electrical cycling stability over thousands of cycles, illustrated in Fig.~\ref{fig_setup}c, where a device is repetitively turned on and off for 3~s and the diffraction is recorded for 1~s during the On stage. If the conductive filament formation is stochastic, each subsequent cycle is uncorrelated. Conversely, if the material has learned a preferential filament and domain configuration, following subsequent cycles are correlated.
\begin{figure}
    \centering
    \includegraphics[width=1\linewidth]{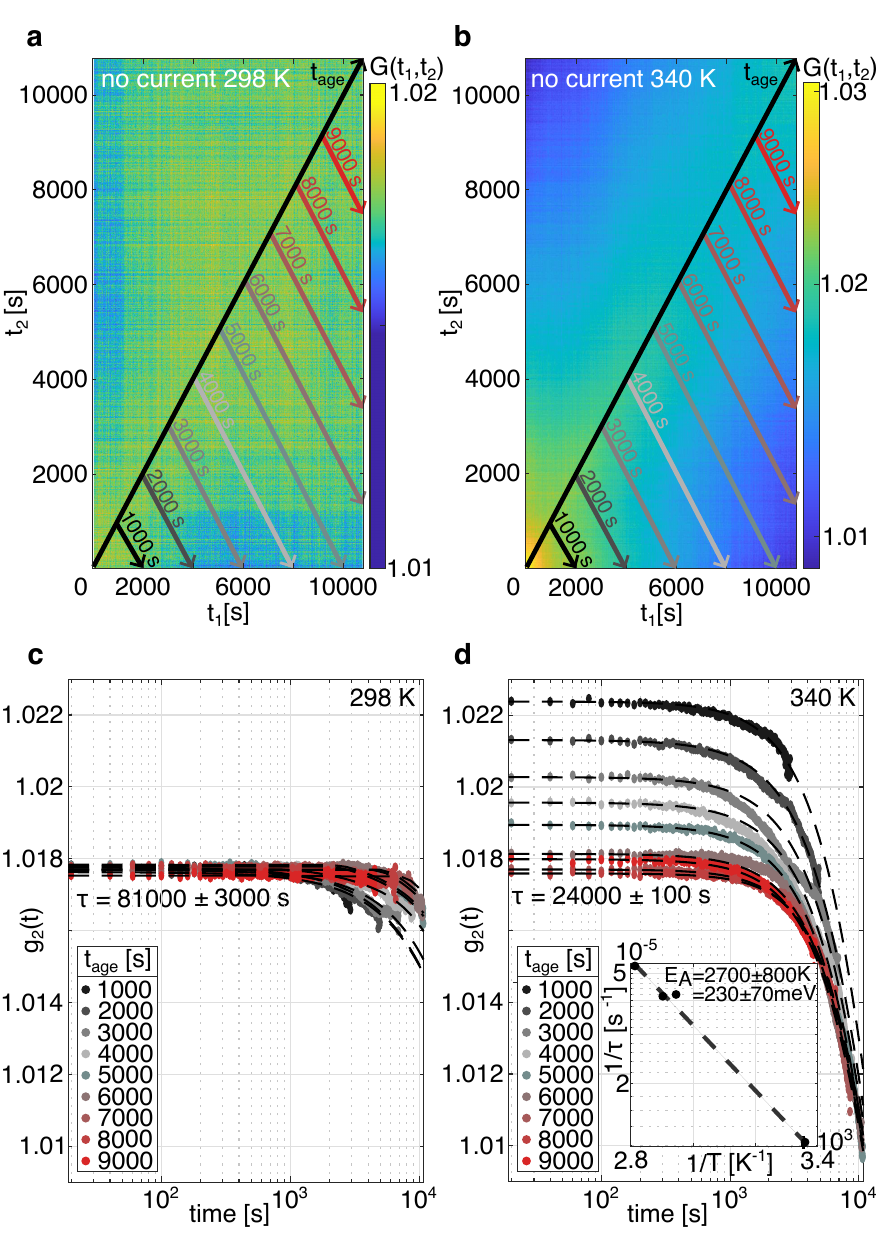}
    \caption{\textbf{Temperature dependence and activation energy.}
    Two-times correlation at room temperature (\textbf{a}) and 340~K (\textbf{b}). The black arrow follows the $t_1 = t_2$ line and represents the aging time $t_\mathrm{age}$. The colored arrows ($t_1 + t_2 = t_\mathrm{age}$) are perpendicular to the black arrow, and correspond to the $g_2(t_\mathrm{age})$ data shown in \textbf{c} and \textbf{d} for various aging times at room temperature and 340~K, respectively. Fits were performed using equation~\ref{eq_exp_decay} with a single $\tau$ value for all $t_\mathrm{age}$. The inset in \textbf{d} shows an Arrhenius plot, giving an activation energy $E_A = 2700 \pm 800$~K ($230 \pm 70$~meV).
    }
    \label{fig_temp}
\end{figure}

\noindent\textbf{Temperature dependence and activation energy in the absence of electric field}\\
Initially, XPCS is measured on VO$_2$ devices as a function of temperature without applying any electric field. Results are shown in Fig.~\ref{fig_temp}. The two-times color plots (Fig.~\ref{fig_temp}a and~\ref{fig_temp}b) show the correlation between diffraction images---and thus microscopic domain distribution--- taken at time $t_1$ and $t_2$. The black diagonal corresponds to an aging time $t_\mathrm{age} = t_1 = t_2$ where the correlation is maximal. Data taken along the colored arrows (perpendicular to the black diagonal, with $t_1 + t_2 = t_\mathrm{age}$) capture the evolution of the domain distribution starting from $t_\mathrm{age}$ and are shown in the lower plots (Fig.~\ref{fig_temp}c and~\ref{fig_temp}d).

The data in the left column, Fig.~\ref{fig_temp}a and~\ref{fig_temp}c, are taken at room temperature (298~K) and provide the slowest observable dynamics in our material or the experimental stability limit, which corresponds to decay rates of 8000$\pm$3000~s. The uniform color in the color plot indicates a stable domain configuration. 
The data in the right column, Fig.~\ref{fig_temp}b and~\ref{fig_temp}d, are taken at 340~K, where thermal energy induces domain reconfiguration and increases the amount of metallic domains (the phase transition is expected around 340~K\cite{zylbersztejn1975metal}).
The decrease in two-times correlation (Fig.~\ref{fig_temp}b) along the $t_\mathrm{age} = t_1 = t_2$ diagonal indicates an unstable regime, where contrast decreases due to the ongoing phase transition in the thermalizing system. 
The average domain size $R_d$ is obtained from the measured Bragg peak's full-width at half-maximum ($Q_\mathrm{FWHM}$) as $R_d = 2\pi/ Q_\mathrm{FWHM}$ yielding $R_d = 20 \pm 4$~nm.
From the $g_2$ correlation decay (Fig.~\ref{fig_temp}c and~\ref{fig_temp}d), the characteristic times $\tau$(T) are obtained at each temperature by fitting the data with equation~\ref{eq_exp_decay}. 
$\tau$(T) follows an activated Arrhenius law, shown in the inset of Fig.~\ref{fig_temp}d, with the associated activation energy $E_a = 230\pm70$~meV, equivalent to $2700\pm800$~K.
Here $E_a$ is comparable to the melting point of VO$_2$, 2240~K, consistent with the view that melting occurs when all domains become freely mobile.
Thus, the temperature dependence confirms that the XPCS experiment is sensitive to domain dynamics in VO$_2$.
\begin{figure*}
    \centering
    \includegraphics[width=1\linewidth]{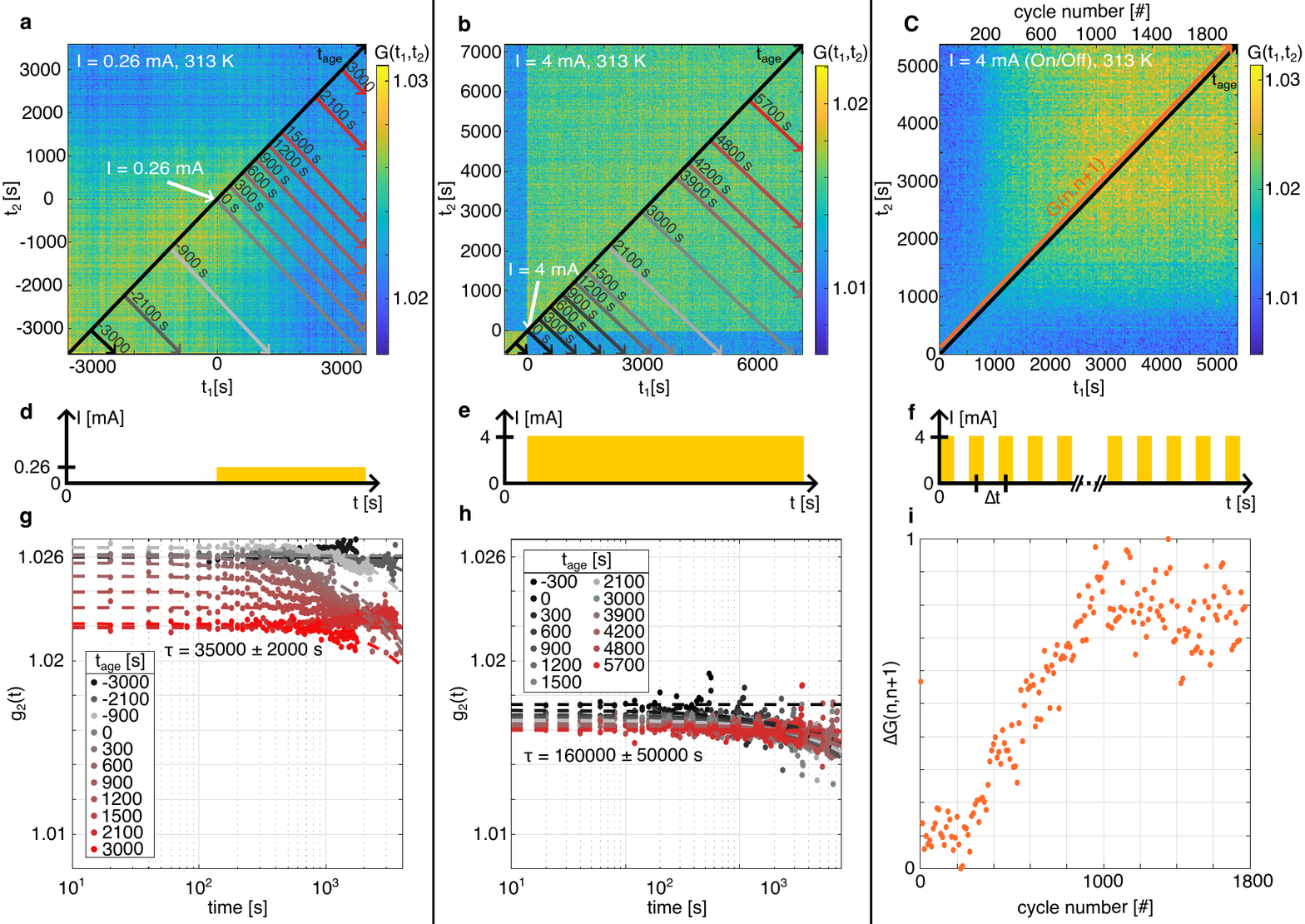}
    \caption{\textbf{Short-term memory, erasure, and learning.}
    \textbf{a} and \textbf{b}, two-times correlation measured as a sub-threshold current I = 0.26~mA and an above-threshold current I = 4~mA is set, respectively. The switching time is set to t = 0~s and the temperature at T = 313.15~K.
    Electrical switching sequences are illustrated in \textbf{d}, \textbf{e}, and \textbf{f}.
    In \textbf{a} the initial state remains unchanged for $\sim$1000~s after switching (memory).
    In \textbf{b} it is immediately erased.
    \textbf{g} and \textbf{h}, $t_\mathrm{age}$ dependent correlation decay $g_2(t)$ obtained from the colored arrows in \textbf{a} and \textbf{b}. Fits were performed using equation~\ref{eq_exp_decay} with a single $\tau$ value for all $t_\mathrm{age}$.
    \textbf{c}, two-times correlation while switching on and off the device every three seconds. The top x-axis shows the cycle number. 
    \textbf{i}, normalized correlation between subsequent switching events $\Delta G(n,n+1)$ as a function of cycle number, corresponding to the orange arrow in \textbf{c}. After 1000 cycles the device has learned a preferred nano-domain configuration resulting in deterministic switching (high correlation).
    }
    \label{fig_switch}
\end{figure*}

\noindent\textbf{Domain dynamics in response to electric fields: short term memory and erasure}\\
XPCS measurements while applying a current are presented in Fig.~\ref{fig_switch}. The switching time is set to 0~s in the time axis.
The temperature was set to 313~K to facilitate switching.
Three switching regimes are shown. First, in Fig.~\ref{fig_switch}a,~\ref{fig_switch}d, and~\ref{fig_switch}g, a sub-threshold current of 0.26~mA is applied, so that no percolating filament forms and the device resistance remains unchanged. 
Before the current is turned on, for t $<$ 0~s, the microscopic domain configuration is stable, indicated by the higher-intensity region (yellow) in the two-times correlation plot (Fig.~\ref{fig_switch}a) and constant $g_2(t)$ in Fig.~\ref{fig_switch}g (data for t $<$ 0~s).
Notably, once the small current is turned on, at t = 0~s (white arrow in Fig.~\ref{fig_switch}a), it takes $\approx$ 1000~s before the initial domain configuration is altered, as indicated by the loss of correlation (blue region in Fig.~\ref{fig_switch}a) and the reduction in $g_2(t)$ (Fig.~\ref{fig_switch}g). This indicates that a small current induces domain reconfiguration leading to slow erasure of the initial state.

When applying an above-threshold current of 4~mA, such that a conducting metallic filament forms across the device, the initial domain configuration is immediately erased (on timescales faster than the 1~s image acquisition rate). However, the newly formed configuration, which includes the metallic filament, is stable for tens of thousands of seconds, as shown in Fig.~\ref{fig_switch}b by the high correlation for all times $t_1$, $t_2$ $>$0~s.
Fig.~\ref{fig_switch}h shows the correlation $g_2(t_\mathrm{age})$ at various aging times before and after switching. The stable On state is characterized by slow domain dynamics with timescales $\tau \approx 150000\pm50000$~s.

\noindent\textbf{Domain configuration upon repetitive switching: learning by plasticity}\\
Figure~\ref{fig_switch}c and~\ref{fig_switch}f show repetitive switching with a high current of 4~mA. The XPCS measurement is synchronized with the switching, such that each subsequent detector image records the diffraction pattern only while the device is switched on. Therefore, the measured correlation indicates how similar the metallic filament and surrounding domain configuration are from one cycle to the next.
The top x-axis in Fig.~\ref{fig_switch}c indicates the cycle number.
For the first $\approx$ 200 cycles, there is little correlation between each On state, as indicated by the low intensity (blue) region in Fig.~\ref{fig_switch}c and low $\Delta G(n,n+1)$ values in Fig.~\ref{fig_switch}i. This shows that initially the filament formation is stochastic.
After $\approx$ 200 cycles the correlation gradually increases and stabilizes for $\approx$ 1000 cycles (Fig.~\ref{fig_switch}i), marking a plastic transition to deterministic filament formation. 
The inset (orange data) shows the normalized correlation between each subsequent cycle, clearly showing that the system reaches a new equilibrium after $\approx$ 1000 cycles.
The device is thus learning a configuration through repetitive switching.

\begin{figure}
    \centering
    \includegraphics[width=1\linewidth]{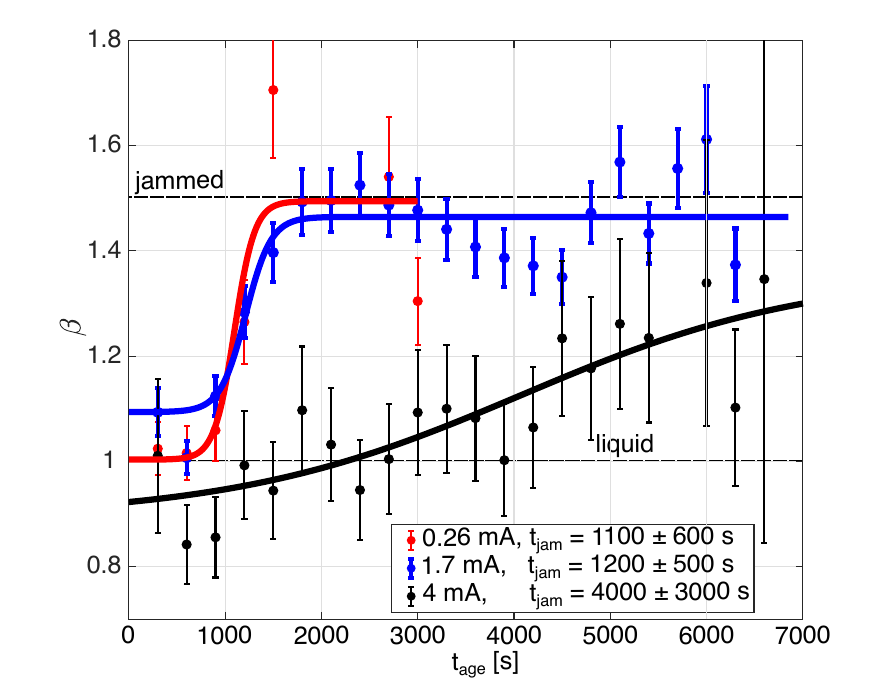}
    \caption{\textbf{Jamming transition.}
    Stretching exponent $\beta$ obtained from fits using equation~\ref{eq_exp_decay} as a function of $t_\mathrm{age}$ for three switching currents.
    $\beta$ = 1 signifies liquid-like dynamics, and $\beta$ = 1.5 jammed/glassy dynamics.
    The solid lines are sigmoid fits, used to extract the transition time $t_\mathrm{jam}$.}
    \label{fig_beta}
\end{figure}

\noindent\textbf{Electrically driven jamming transition}\\
The correlation decay of $g_2(t)$ in Fig.~\ref{fig_switch} is fitted using equation~\ref{eq_exp_decay} to extract the characteristic timescales $\tau$ and the stretching exponent $\beta$. The exponent $\beta$ characterizes the type of observed dynamics. 
$\beta$ = 1 indicates a continuous, unconstrained, and liquid-like dynamics. 
$\beta$ = 1.5 describes a non-diffusive, constrained, and glassy/jammed dynamics\cite{shpyrko2014x}.
The value of $\beta$ as a function of $t_\mathrm{age}$ after electrical switching is shown in Fig.~\ref{fig_beta} for three switching currents, 0.27, 1.7, and 4~mA. For all currents a transition from $\beta$ = 1 to $\beta$ = 1.5 is observed, \textit{i.e.}, a jamming transition is driven by a fixed current.
Remarkably, while jamming is expected in soft matter and granular systems, here it is observed in hard matter, implying granular-like dynamics of the nano-domains.
The data are fitted with a sigmoid function to determine a transition time $t_\mathrm{jam}$, indicated in the legend.
Notably, the larger the current, the slower the transition, reminiscent of shear-thickening in non-Newtonian fluids and glasses.
\\

\noindent\textbf{Discussion}\\
The Bragg XPCS data presented here tracks changes in the microscopic nanoscale arrangements of metallic and insulating domains in VO$_2$ neuromorphic devices, both as a function of temperature (Fig.~\ref{fig_temp}) and current (Fig.~\ref{fig_switch}). While XPCS does not give real-space information other than average domain size and dynamics timescales, it characterizes the nanoscale distribution within the micron-sized beam in a single shot.
The measurements detect changes on a 1~s timescale over periods of hours, in contrast to real-space nano-phase imaging techniques, such as nano-diffraction\cite{shabalin2020nanoscale,rippy2019,smith2025nanodiffraction,d2025self}, fluorescence\cite{DAnna_Xray_fluorescence}, dark-field X-ray microscopy\cite{salev2024local,kisiel2025high}, and reflectivity\cite{del2021spatiotemporal}, which are typically slower. Faster techniques, such as electrical measurements, are mostly ensemble measurements not sensitive to single domains\cite{mangu2025hidden}. 
The temperature dependence shows that domain reconfiguration is thermally activated, with an activation energy $E_a = 230\pm70$~meV ($E_a = 2700\pm800$~K), which is close to the VO$_2$ melting temperature (2240~K).
Interestingly, when applying current to a device, the observed domain dynamics timescales are comparable to the thermally activated timescales, measured in tens of thousands of seconds, suggesting that the dynamics may be induced by Joule heating\cite{kalcheim2020non,del2021spatiotemporal}.
These are slow processes that happen after the initial formation of a metallic filament. By changing the initial nano-domain configuration, they directly impact the amount of randomness in filament formation.

Electrical switching induces a jamming transition from liquid-like to glassy dynamics (Fig.~\ref{fig_beta}). The transition occurs on timescales of thousands of seconds. This may be due to two effects. The number of metallic domains may increase, resulting in a reduction of space available for unconstrained rearrangement, as observed in temperature-dependent nano-diffraction imaging experiments\cite{d2025self,shao2025x}. Alternatively, the domains may all be moving in a preferential direction, for example towards the conductive filament, effectively reducing the available space between domains.
Similar glassy behaviors were observed with XPCS in the phase-change metal-oxide LSCO\cite{danna2026_LSCO}.
The jamming transition time $t_\mathrm{jam}$ is lower for lower switching currents, indicating that domains reach a packed arrangement faster with smaller currents. If the dynamics are activated by thermal energy from Joule heating, it stands to reason that the larger currents give larger phase-space for the domains to explore, and so a longer time until they reach a restricted distribution.

The devices are observed to have a controllable short-term memory encoded in the microscopic domain configuration.
While the application of a large current that forms a metallic filament immediately erases the memory (on timescales $<$1~s) contained in the domain configuration (Fig.~\ref{fig_switch}b and~\ref{fig_switch}h), a small sub-threshold current slowly changes the configuration over thousands of seconds (Fig.~\ref{fig_switch}a and~\ref{fig_switch}g).
The electrically induced transition is believed to be generally thermally driven through Joule heating\cite{kalcheim2020non,del2021spatiotemporal,cheng2025purely}. Therefore, the initial domain configuration is likely to control where current flow is large and thus where more heating occurs and ultimately where the metallic filament forms.
Thus, slow erasure of the initial state by a sub-threshold current has the potential to increase the stochasticity of filament formation, in particular if applied between two above-threshold current pulses.
Similarly, the observed fast erasure of the initial state upon switching with a large current, as indicated by the lack of correlation between data at positive and negative times in Fig.~\ref{fig_switch}b, is responsible for stochastic filament formation upon electrical cycling.

The amount of stochasticity in electrical switching is likely related to the ability of the material to return to its initial state at the end of a switching cycle (when the current is turned off). 
Forming the metallic filament immediately ($<$1~s) erases the initial domain configuration information, however it is possible to take advantage of the jamming transition that occurs on much longer timescales ($\sim$~1000~s) to induce long-term memory through plasticity. This is observed upon repetitive switching with a high current (4~mA, Fig.~\ref{fig_switch}c and~\ref{fig_switch}i), where the filament formation is initially stochastic (low-correlation region at t $<$ 1000~s and cycle number \mbox{$<$ 200}) and gradually becomes deterministic after the system is driven into the glassy state, as indicated by the high-correlation region. 
Thus, by driving the device into the glassy state, it becomes sufficiently plastic to remember its previous switched state, effectively enabling training. These results are corroborated by electrical measurements of nanodevices subjected to voltage pulses\cite{del2019subthreshold}, where persistent metallic domains are found to act as long lived memory and reduce switching stochasticity. Similarly, scanning near-field optical microscopy revealed metallic patches that persist after high currents\cite{tiwari2026near}.
Combined with the observed erasure of the initial domain configuration by a sub-threshold current, this gives the ability to train and reset VO$_2$ neuromorphic devices.

In conclusion, coherent Bragg scattering was used as a method to track changes in the nanoscale domain distribution throughout VO$_2$ neuromorphic devices under operation. The results show that the exact domain distribution before electrical switching influences the metallic filament formation and that the initial distribution controls the stochasticity of the filament formation.
In particular, it is found that the initial domain configuration can be erased by a sub-threshold current, or driven into a jammed/glassy state by repetitive above-threshold currents, which together allow stochastic and deterministic switching. The jamming transition, as well as the short-term erasure, occur on slow timescales of thousands of seconds, justifying the need to look at nanoscale information over long timescales in phase-change materials.
Regarding neuromorphic computing and emulating neural networks, this gives a possible path forward: a neural network where the connections (synapses) are made with VO$_2$ may be plastically trained by inducing the jamming transition through repetitive switching, and it may be erased by applying a low sub-threshold current.

\noindent\textbf{Methods}\\
\textit{Sample Preparation:} 
45~nm thick VO$_2$ films were grown on (1 -1 0 2)-oriented R-cut Al$_2$O$_3$ substrates via reactive RF magnetron sputtering using a V$_2$O$_3$ target in a 4~mTorr Ar/O$_2$ (93/7\%) atmosphere and at 470~$^{\circ}$C substrate temperature. After the growth, the samples were maintained in the same atmosphere and cooled at a rate of 12~$^{\circ}$C min$^{-1}$ to help preserve the correct oxygen stoichiometry.
(100~nm~Au)/(20~nm~Ti) electrodes were fabricated in a 5~$\mu$m $\times$ 5~$\mu$m two-terminal geometry using standard photolithography techniques and e-beam evaporation. After the as-grown sample had been characterized, devices were globally irradiated using a 30~keV focused Ga$^+$ ion beam with a fluence of 6.24$\cdot$10$^{13}$ Ga ions/cm$^2$ using a commercial scanning electron microscope.

\textit{XPCS setup:}
The \mbox{X-ray} photon correlation spectroscopy experiments were conducted at the 9C beamline of the Pohang Light Source-II (PLS-II)\cite{ham2026upgrade}. 
The devices were placed into a custom-built four-probe station\cite{ham2022multimodal} connected to a Keithley SourceMeter 2612B.
The 6~keV \mbox{X-ray} beam was focused with KB mirrors to a spot size of $7.4\times14.8~\mu$m$^2$ and a transverse coherence length of 1.7~$\mu$m in both directions. The photon flux was $2\times10^{10}$~ph/s.
The diffraction was recorded with a \mbox{EIGER2 X 1M} detector, with 1028$\times$1062 pixels each 75~$\mu$m in size. The detector was located 861.29~mm from the sample at a 2$\theta$ angle of 50$^\circ$.

\textit{XPCS analysis:}
XPCS data are analyzed by correlating 2D diffraction images in time.
The two-time intensity-intensity correlation $G(t_1, t_2)$ correlates the diffraction at $t_1$ and $t_2$, which gives three-dimensional plots such as in Fig.~\ref{fig_switch}a,~\ref{fig_switch}b, and~\ref{fig_switch}c.
$G(t_1, t_2)$ is calculated as  
\begin{equation}
    G(t_1, t_2) = \frac{I(t_1)  I(t_2)}{\langle I^2 \rangle_{t}},
\label{eq_2times}    
\end{equation}
where $\langle I^2\rangle_{t}$ denotes the average of the squared intensity over the total acquisition period\cite{shpyrko2014x}.
Averaging values of $G(t_1, t_2)$ with constant time delays $t_2 - t_1 = t$ leads to the one-time correlation functions
\begin{equation}
    g_2(t) = \frac{\langle I(\tau)  I(\tau+t)\rangle_{\tau}}{\langle I^2 \rangle_{\tau}}.
\label{eq_g2}    
\end{equation}

The $t_1 = t_2$ line corresponds to the aging time $t_{age}$ (black arrow in Fig.~\ref{fig_temp} and~\ref{fig_switch}). 
Lines perpendicular to $t_1 = t_2$ starting at $t_1 = t_2 = t_\mathrm{age}$ (colored arrows in Fig.~\ref{fig_temp} and~\ref{fig_switch}, $t_1 + t_2 = t_\mathrm{age}$) give the one-time correlation functions $g_2^{t_\mathrm{age}}(t)$.
The intermediate scattering function  $|F(t)|$ is calculated using $g_2(t) = 1 + A|F(t)|^2$, where $A$ describes the beam coherence\cite{sutton1991observation}.
The correlation decay is fit with a single exponential decay function,
\begin{equation}
|F_{t_\mathrm{age}}(t)| = \exp(-(t/\tau)^\beta),
\label{eq_exp_decay}
\end{equation}
where $\tau$ is the characteristic relaxation time and $\beta$ the stretching exponent.
Note that in this work, a single $\tau$ value is fit for all $g_2^{t_\mathrm{age}}(t)$ from each XPCS measurement.

\textit{Electrical Measurements:}
Electrical transport measurements were performed in the four-probe station used for the XPCS measurements with a Keithley source meter 2612B. 
The results in Fig.~\ref{fig_RT} are from the samples studied here by XPCS.
Temperature dependence (Fig.~\ref{fig_RT}a) was measured with a 1~$\mu$A current. Thermal switching and associated hysteresis confirm expected device function\cite{ghazikhanian2025}.
Electrical switching (Fig.~\ref{fig_RT}b) was characterized right before the XPCS experiments, to decide switching currents. The currents $I_1$ and $I_2$ in Fig.~\ref{fig_RT}b indicate multiple switching events where a conductive filament forms. Thus, at 4~mA used during the switching XPCS experiment (Fig.~\ref{fig_switch}) the measured device region is fully switched to the On state.

\begin{acknowledgments}
We acknowledge the Pohang Accelerator Laboratory (PAL) for provision of synchrotron radiation facilities and we would like to thank Su Yong Lee and Daseul Ham for assistance and support in using beamline 9C. 
This research was supported by the Quantum Materials for Energy Efficient Neuromorphic Computing (Q-MEEN-C), an Energy Frontier Research Center funded by the US Department of Energy (DOE), Office of Science, Basic Energy Sciences, under Award DE-SC0019273.
This work was performed in part at the San Diego Nanotechnology Infrastructure (SDNI) of UCSD, a member of the National Nanotechnology Coordinated Infrastructure, which is supported by the National Science Foundation (Grant ECCS-2025752).
\end{acknowledgments}

\bibliographystyle{naturemag}
\bibliography{biblio}	

\end{document}